\documentclass{aastex631}

\usepackage{longtable}
\usepackage{multirow}
\usepackage{subfigure}
\usepackage{tablefootnote}
\usepackage{ulem}

\begin{document}

\title{The Temporal Symmetrical and Translational Structure in Gamma-Ray Burst Light Curves}

\author{Dong-Jie Liu}
\affiliation{Department of Astronomy, School of Physics, Huazhong University of Science and Technology, Wuhan, 430074, China}

\author[0000-0002-5400-3261]{Yuan-Chuan Zou}
\affiliation{Department of Astronomy, School of Physics, Huazhong University of Science and Technology, Wuhan, 430074, China}
\affiliation{Purple Mountain Observatory, Chinese Academy of Sciences, Nanjing, 210023, China}

\begin{abstract}
Tremendous information is hidden in the light curve of a gamma-ray burst (GRB). Based on CGRO/BATSE data, \cite{2021ApJ...919...37H} found a majority of GRBs can be characterized by a smooth, single-peaked component superposed with a temporally symmetrical residual structure, i.e., a mirror feature for the fast varying component. In this study, we conduct a similar analysis on the same data, as well as on Fermi/GBM data. We got a similar conclusion that most GRBs have this symmetrical fast varying component. Further more, we chose an alternative model to characterize the smooth component and used a three-parameter model to identify the residual, i.e., the fast component. By choosing 226 BATSE GRBs based on a few criteria, we checked the time symmetrical feature and time translational feature for the fast components and found the ratio is roughly 1:1. We propose that both features could come from the structure of the ejected shells. Future SKA might be able to observe the early radio emission from the collision of the shells.
\end{abstract}


\section{Introduction} \label{sec:1}

Gamma-ray bursts (GRBs) are among the most powerful explosions in the universe, with duration ranging from less than 0.01 seconds to more than 1000 seconds. The isotropic energy emitted in $\gamma$-rays can be as high as $10^{48}-10^{55}$ ergs. GRBs are observed to consist of two different phases: the prompt emission, which is concentrated in the keV-MeV energy range, and the afterglow, which extends from $\gamma$-rays to the radio band \citep{2006RPPh...69.2259M,2015PhR...561....1K}. According to the fireball model, GRBs are produced by highly relativistic and collimated jets, and the prompt emission is the result of the kinetic energy dissipated in internal collisions, while the afterglow is produced by the interaction of the jet with the ambient material \citep{1999PhR...314..575P,2004RvMP...76.1143P}.

As the nature of GRBs is not completely understood, studying their light curves may provide useful information. In this paper, we focus on the light curve of prompt emission. Most GRB light curves are highly variable, with variability timescales that can be as short as milliseconds \citep{2012MNRAS.425L..32M}. However, there are also about $20\%$ of bursts that show a smooth pulse, which can generally be fitted by a fast rise exponential decay (FRED) model \citep{1996ApJ...459..393N,2000ApJS..131....1L,2000ApJS..131...21L,2003ApJ...596..389K}. According to the internal-shock model, a GRB light curve is composed of many individual pulses, with every individual pulse corresponding to a collision between individual shells \citep{1996ApJ...459..393N,1997ApJ...490...92K}. However, there are also models and studies believe that GRB pulses can be made up of two components, a smooth component and a more variable component. \cite{2006A&A...447..499V} considered the two components as a slow component and a fast component and found that the slow component is generally softer than the fast ones. \cite{2011ApJ...726...90Z} proposed the Internal-Collision-induced MAgnetic Reconnection and Turbulence (ICMART) model, which also prefers the two-component scenario. \cite{2012ApJ...748..134G} developed a new method to identify significant clustering structures of a light curve in the frequency domain and found that the majority of bursts have clear evidence of such a superposition effect.

In a study of a dataset of GRBs from the Burst And Transient Source Experiment (BATSE) at NASA's Compton Gamma Ray Observatory \citep{1993ApJ...413..281B}, \cite{2021ApJ...919...37H} found that a majority of GRB pulses could be characterized by a smooth, single-peaked component coupled with a temporally symmetrical residual structure. This finding is intriguing as it provides further evidence that GRB light curves may indeed have two components. Additionally, the temporal symmetry of the residual structure is unexpected and has not been predicted by any existing model to our knowledge. Very recently, \cite{2022arXiv220807236M} observed the time reflection effect in laboratory. It could be a clue for understanding the temporally symmetrical effect in GRB light curves.

We aim to investigate the existence of the symmetrical signal and determine its nature. In this study, we follow the method employed by \cite{2021ApJ...919...37H} to confirm the time symmetry of the residual and expand the data set by using data from the Gamma-ray Burst Monitor (GBM) aboard the Fermi Gamma-ray Space Telescope \citep{2009ApJ...702..791M}. We also modify the method to mitigate the influence of the slow component and propose a three-parameter model to characterize the residual. Finally, we attempt to interpret this phenomenon using the internal-external shock model.

In this paper, we first introduce the method in \cite{2021ApJ...919...37H} for characterizing a GRB light curve, and present our results based on our data selection in Section \ref{sec:2}. In Section \ref{sec:3}, we discuss limitations of the original model and propose a new model to further investigate this symmetry. Section \ref{sec:4} presents the findings of our new model. We also propose a possible explanation for the phenomenon in Section \ref{sec:5}. Finally, we conclude and discuss our findings in Section \ref{sec:6}.  

\section{The original model and results} \label{sec:2}

\subsection{Method} \label{sec:2.1}

We will briefly introduce the method used in \cite{2021ApJ...919...37H} first. For a given GRB light curve, the monotonic component of the GRB pulse can be modeled by fitting the light curve using a simple, generic mathematical model. The pulse model is based on the pulse intensity function of \cite{2005ApJ...627..324N}, which can be described as 
\begin{equation}
    I(t)=A\lambda e^{[-\tau_1/(t-t_s)-(t-t_s)/\tau_2]}, 
\end{equation}
where $t$ is the time since trigger, $A$ is the amplitude of the pulse, $t_s$ is the pulse start time, $\tau_1$ is the pulse rise parameter, $\tau_2$ is the pulse decay parameter, and $\lambda$ is a normalization constant. Since not all GRB pulse light curves can be fitted by this function, the remaining light curves are fitted using a Gaussian distribution function of the form
\begin{equation}
    I(t)=\frac{C}{\sigma\sqrt{2\pi}}{\rm exp}[-(\frac{t-t_0}{\sqrt{2}\sigma})^2], 
\end{equation}
where $C$ is the pulse amplitude, $t_0$ is the time when $C$ occurs, and $\sigma^2$ is the variance. The pulse duration window can be defined by the pulse starting time $t_{start}$ and the pulse end time $t_{end}$, both of which are measured at $I_{meas}/I_{peak}=e^{-3}$. A background model is also required with a simple form
\begin{equation} 
    B=B_0+BS\times{t}, 
\end{equation}
where $B_0$ is the mean background, and $BS$ is the rate of change of the background. After fitting the monotonic component, the residual, which are considered to be temporally symmetric, can be obtained by subtracting this component from the data. Two parameters are required to characterize the residual: the time $t_{0;mirror}$ at which the forward and backward residuals are symmetrical and the stretching parameter $s_{mirror}$, which represents the ratio between time-forward structures and time-reversed structures. Both of these parameters are determined by the maximum value of the normalized cross-correlation function (CCF):
\begin{equation}
    {\rm CCF}=\frac{\sum_i (x_i-\overline x)(y_i-\overline y)}{\sqrt{\sum_i (x_i-\overline x)^2} \sqrt{\sum_i (y_i-\overline y)^2}}. 
\end{equation}
In other words, after obtaining the residual, then the CCF is applied to the folded time-forward part and time-reversed parts with varying $t_{0;mirror}$ and $s_{mirror}$ values to search for the maximum CCF value, which indicates the best match between the time-forward and time-reversed parts. Note that the existence of $s_{mirror}$ results in different bin widths between the two parts, linear interpolation is employed on the folded part of the residual light curve. Data resampling is used for error estimation \citep{2010arXiv1009.2755A}. Different $t_{0;mirror}$ and $s_{mirror}$ are generated by adding random Poisson noise to the light curve. The distribution of parameters are used to evaluate their uncertainties. A GRB light curve with $\sigma_{s,mirror}<0.4$ can be directly classified as temporally symmetric. Besides, a residual statistic, given by  
\begin{equation}
    R = \sqrt{\frac{\sigma_{res}}{\langle B \rangle}}, 
\end{equation}
is introduced to quantify the brightness of the residual structure. Additionally, the $p$-value obtained from a $\chi^2$ test on the pulse fitting is required to assess the quality of the fitting result.

For a more detailed description and the criterion of temporally symmetry, please refer to \cite{2014ApJ...783...88H}, \cite{2018ApJ...863...77H} and \cite{2021ApJ...919...37H}. 

\begin{figure}[t!]
\epsscale{0.85} \plotone{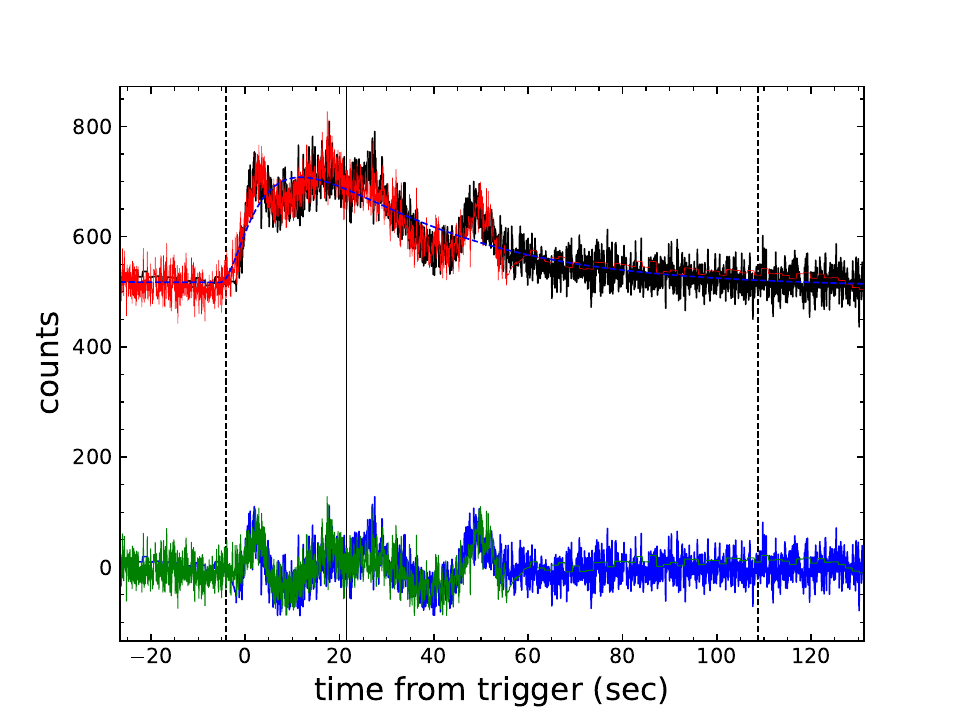}
\caption{Temporally symmetric model fits to BATSE pulse 659 light curve. The figure shows the counts data in black line, the fit to the monotonic component in blue dashed line, the time-reversed model in red line, the residual in blue line, the time-reversed residual in green line, the duration window from $t_{start}$ to $t_{end}$ (vertical dashed lines), and the time of reflection $t_{0;mirror}$ (vertical solid line).
\label{fig:1}}
\end{figure}

A typical fitting result is presented in Figure \ref{fig:1}, 
which shows the model applying to BATSE pulse 659. The fitting looks well. The residual structure shows time symmetry with the maximum CCF of residual ${\rm CCF}_{resids}=0.468$ and $t_{0;mirror}=21.49$, $s_{mirror}=0.69$. By resampling the data and using Monte Carlo simulation, a seires of $t_{0;mirror}$ and $s_{mirror}$ can be generated for error estimation. For BATSE pulse 659, a $\sigma_{s,mirror}$ of 0.14 is obtained, indicating that it is a temporally symmetrical light curve.

\subsection{Data selection} \label{sec:2.2}

In this work, we analyze two different GRB data samples obtained from BATSE and GBM, respectively. The following section provides a brief description of each data set.

We repeat Hakkila's results using the same BATSE data set as in \cite{2021ApJ...919...37H}, but conduct data filtering. Our data comes exclusively from BATSE's 64 ms resolution data \footnote{\url{https://heasarc.gsfc.nasa.gov/FTP/compton/data/batse/ascii_data/64ms/}}, and we do not attempt to analyze the 4 ms data, since the proportion of such data is very small in the original data set and would not significantly affect the results.  We remove data considered to be multiple pulses in \cite{2021ApJ...919...37H}. 
These GRBs are believed to have multiple emission episodes, which make it inappropriate to treat them as a single GRB. Therefore, it is necessary to segment the data. Since we do not have the specific parameters used by the original authors for segmentation, we decided to discard these data. These data may have no much impact on the final results or proportions. Our sample includes 226 GRBs out of the 312 BATSE GRBs, allowing us to replicate previous results.

We further consider Fermi GBM data of the years 2020 and 2021. We downloaded the time-tagged event (TTE) data for approximately 600 GRBs from the Fermi Science Support Center (FSSC) FTP website \footnote{\url{https://heasarc.gsfc.nasa.gov/FTP/fermi/data/gbm/bursts/}}. GBM comprises 12 sodium iodide (NaI) detectors and 2 bismuth germanate (BGO) detectors. For each GRB, we select the TTE data from the triggered NaI detectors since they are typically the brightest ones. Subsequently, we combine the data from these detectors to extract the 64 ms light curve data. The data extraction process is performed using the \emph{GBM Data Tools} \footnote{\url{https://fermi.gsfc.nasa.gov/ssc/data/analysis/gbm/}}, a software package provided by the FSSC.

Although the energy ranges and time scales of the data from the BATSE and GBM instruments differ, considering that our research focus is on the overall morphology of the light curves, the impact of energy ranges and time scales is not significant.

\subsection{Results}

By following the procedures outlined in Section \ref{sec:2.1}, GRBs can be categorized into three groups depending on their conformity to the temporally symmetric model. Utilizing the data sample detailed in Section \ref{sec:2.2}, we present our statistical outcomes in Table \ref{tab:1}. To facilitate comparison, we list the findings of \cite{2021ApJ...919...37H} in the initial row. The data selected from \cite{2021ApJ...919...37H} were unbiased. To demonstrate this, we have listed the same data distribution in \cite{2021ApJ...919...37H} in the second row. The symmetry ratio of the new sample is nearly consistent with that of the original sample. Followed by our outcomes acquired from the identical BATSE data and new data from GBM.

Our findings reveal that the temporally symmetric model was able to successfully fit $85\%$ of the BATSE GRBs, which is highly consistent with the previous results reported by \cite{2021ApJ...919...37H}. However, when using the data obtained from GBM, we observed a significant difference in the distribution of results, with the majority of the data being classified as monotonic. This difference could be attributed to the fact that the effective detection area corresponding to the GBM data in our sample is considerably smaller than that of the BATSE, resulting in a lower signal-to-noise ratio of the majority of the GRBs detected by GBM and less distinct structures, which lead to a smaller residual statistic $R$. Nonetheless, the proportion of monotonic GRBs does not affect the symmetry ratio. Among the remaining GRBs, the proportion is over $70\%$, which although lower than before, is still relatively high.

The fitting parameters and corresponding results for the data from both instruments are presented in Tables \ref{tab:2} and \ref{tab:3} in the appendix. It should be noted that for the GBM data, we only report the results for the other two types of pulses as there were a large number of monotonic pulses identified that do not contribute to the determination of the symmetry ratio. It is worth mentioning that the absence of pre-screening of the GBM data may introduce biases, which could potentially be a contributing factor to the lower symmetry ratio observed in the GBM data.
\begin{deluxetable*}{lcccc}
\tablewidth{0pt} 
\tablecaption{The Results of Temporal Symmetry Pulse Fitting \label{tab:1}}
\tablehead{
\colhead{Data set} & \colhead{Not temporally symmetric\tablenotemark{a}} & \colhead{Monotonic\tablenotemark{b}} & \colhead{Temporally symmetric\tablenotemark{c}} & \colhead{Symmetry ratio}
}
\startdata
BATSE \citep{2021ApJ...919...37H}& 25& 126& 161& $86.6\%$\\
BATSE with filter & 21& 83& 122& $85.3\%$\\
\hline
BATSE& 18& 106& 102& $85.0\%$\\
GBM 64 ms (2020)& 14& 195& 39& $73.6\%$\\
GBM 64 ms (2021)& 11& 170& 37& $77.1\%$\\
\enddata
\tablecomments{
\tablenotetext{a}{\scriptsize The pulse cannot fit or inconsistent with model ($p < 0.05$ and $R \geqslant 2.0$ and $\sigma_{s,mirror} \geqslant 0.4$).}
\tablenotetext{b}{\scriptsize The pulse fitted by a monotonic pulse model but with a weak residual structure ($p > 0.05$ or ($p < 0.05$ and $R < 2.0$)).}
\tablenotetext{c}{\scriptsize The residual structure is found to be temporally symmetrical ($p < 0.05$ and $\sigma_{s,mirror} < 0.4$).}
}
\end{deluxetable*}

\section{A new model} \label{sec:3}
\begin{figure}[t!]
\plottwo{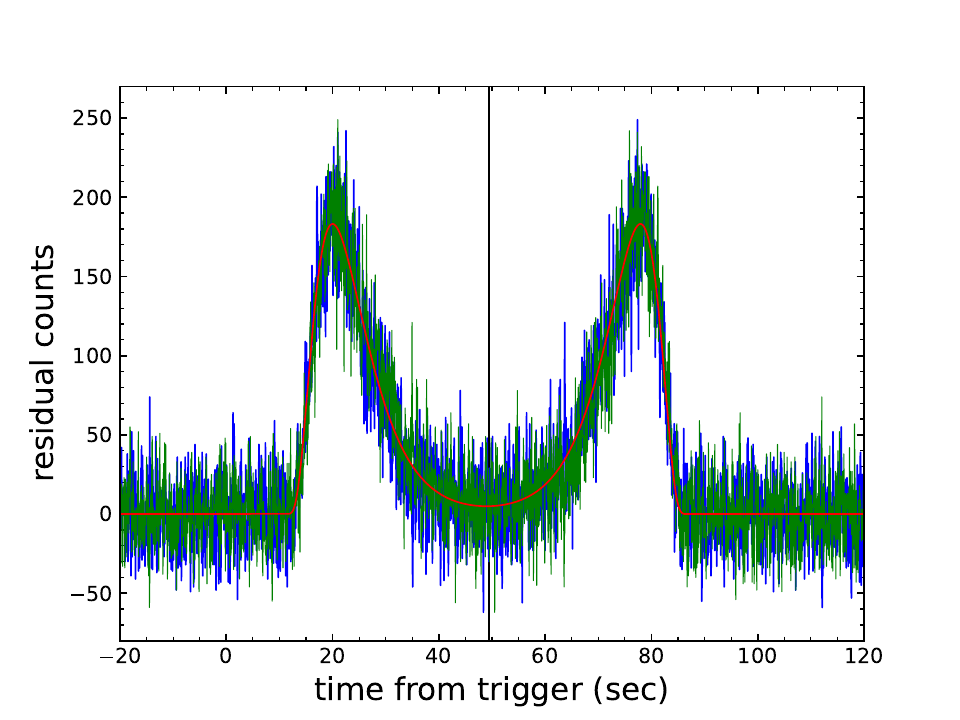}{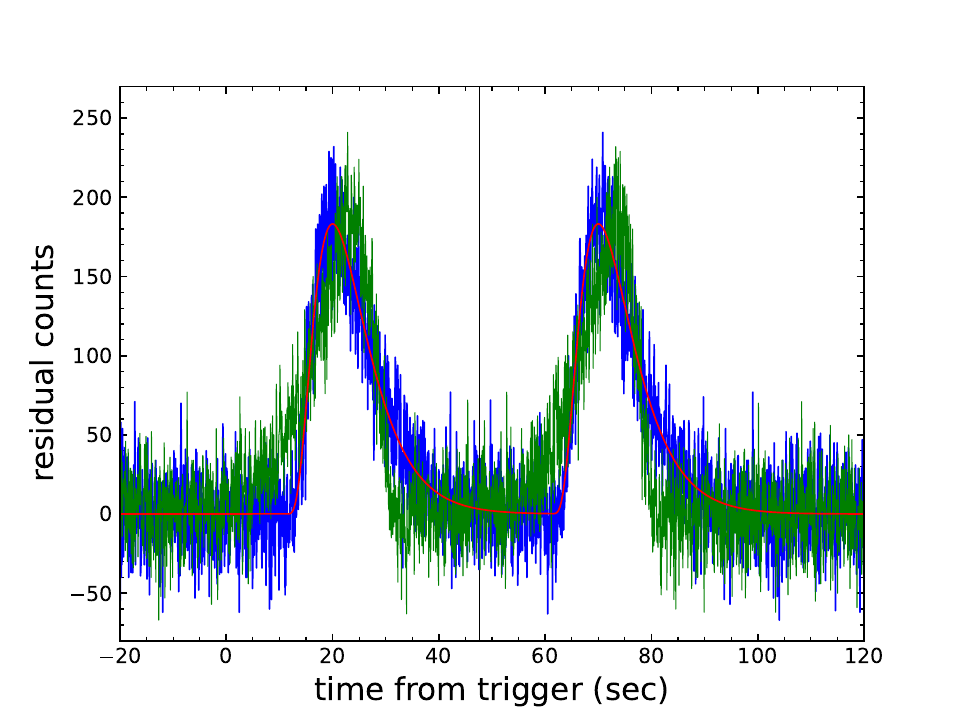}
\caption{Synthetic residuals and fitting results. Including the signal in red, the signal with noise in blue, the time-reversed residuals in green, the time-symmetrical residual in the left panel and the time-translational residual in the right panel. 
\label{fig:2}}
\end{figure}

Another issue highlighted in \cite{2021ApJ...919...37H} is that the assumption of temporal symmetry may not fully capture the shape of residual structures. This can be seen in Figure \ref{fig:2}, where we manually generate two types of residuals corresponding to the time-symmetrical and time-translational cases. The red curve represents the original data, which we then perturb with Poisson noise to obtain the blue curve. The green curve is the time-reversed residual determined by the maximum value of CCF. Through Monte Carlo simulations, we find that the calculated value of $\sigma_{s,mirror}$ is small enough in both cases to be defined as temporally symmetric. This means that the original model can only characterize the symmetry of pulse orders in the fast-varying structure but cannot capture the symmetry of the pulse shapes. Therefore, building upon the original model, we design a new model that characterizes the residuals. Under this model, we can provide a comparison of pulse shape symmetry and translation. For convenience, we refer to these models as the symmetrical model and translational model, respectively. We expect the symmetrical model outperforms the translational model, which indicates that symmetry exists not only in the pulse order but also in the pulse shapes.

The details of the new model are described in the following subsections.
\subsection{Characterizing the monotonic component} \label{sec:3.1}
Several studies have proposed models that describe GRB pulses as a combination of slow and fast components \citep{2006A&A...447..499V,2011ApJ...726...90Z,2012ApJ...748..134G}. This aligns closely with the viewpoint of \cite{2021ApJ...919...37H} that GRBs can be characterized by a smooth, single-peaked component superposed with a temporally symmetrical residual structure. By combining the two, it naturally leads us to hypothesize that the fitting to the monotonic component represents the slow component and the residual represent the fast component. Additionally, the unique properties of the fast component can distinguish it from the slow component.

\begin{figure}[ht!]
\plottwo{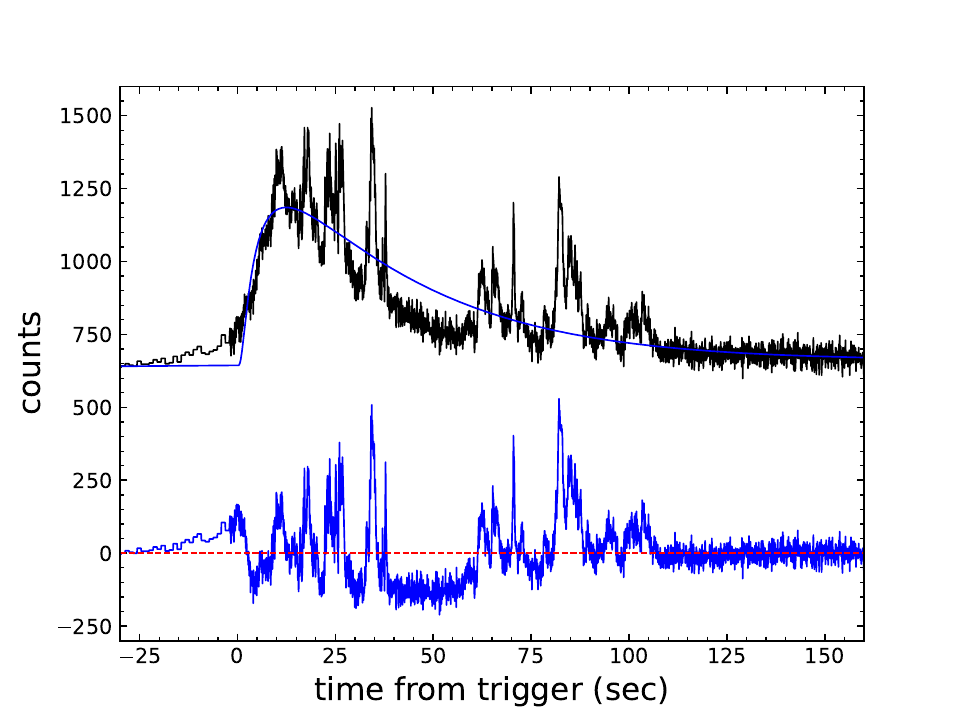}{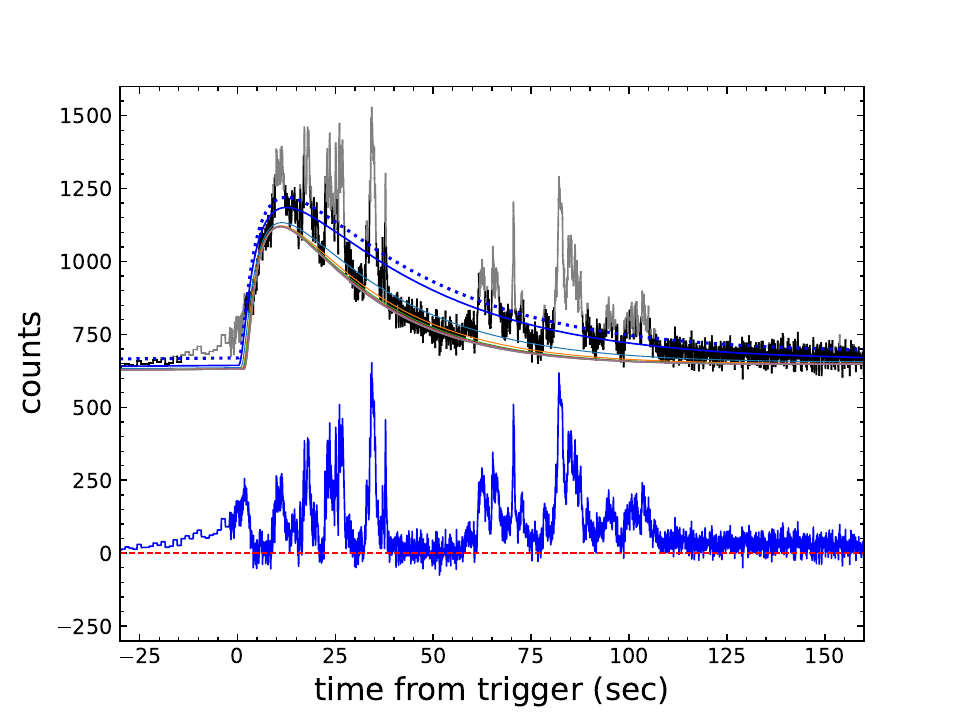}
\caption{Two fitting approaches apply to BATSE pulse 109, a direct fitting (left panel) which produce the undesirable residual and the iterative fitting (right panel). The figure shows the counts data in black line, the fit to the Norris/Gaussian model in colored solid lines, the residual in blue line, and a zero horizontal line in red dashed line.
\label{fig:3}}
\end{figure}

If both the fast component and the slow component are from some certain radiation processes, i.e., not absorption as the non-thermal spectra in most cases, we do not expect any of them are significantly below zero.
However, a direct fitting to the GRB light curve can leave residual component that are obviously below zero. The left panel of Figure \ref{fig:3} shows a direct fitting to BATSE pulse 109. According to our hypothesis, the monotonic component should be lower than the fitting result, but due to the existence of the fast component, the fitting result is raised, leaving a residual that does not meet our expectations. In other words, due to the presence of fast-varying components, the fitting of slow-varying components can be overestimated, leading to an underestimation of the fast-varying components in the results. Representing the slow-varying component solely based on direct fitting results is inadequate.

To address this problem, we have developed a simple yet effective method. We continue to use the FRED function in \cite{2005ApJ...627..324N} or Gaussian distribution function as the basis of the pulse model. The difference is that we will perform iterative fitting. After each fitting, the data points that are above the noise level of the fitting curve will be masked and the rest of the data will be used for the next fitting, until the fitting results converge. The parameter results of the last fitting are taken as the initial value of the next fitting. The noise level is estimated simply by taking the square root of the signal. In the right panel of Figure \ref{fig:3}. Initially, we perform a fitting on the entire light curve, resulting in the blue solid line. Using this blue solid line, we obtain the critical line, represented by the blue dotted line in the graph. The data points above the critical line, represented by the gray region, will be removed. Then the remaining data points are used for the second round of fitting and critical line assessment. As the iteration progress, the number of removed data points gradually decrease, and eventually, the fitting line remain relatively stable, indicating convergence. For BATSE pulse 109, the final fitting curve converges to the expected contour after 9 iterations. The valleys of residual show Poisson variations near zero. Notice that the iterative fitting is the main difference from Hakkila's method. The consequence is that the negative part in the light curve never appear in our treatment. This treatment implies an assumption that there is no absorption, and all the emissions are from optically thin regions. Using this method, we obtain the fitting results of the monotonic/slow component of GRBs, which are shown in Table \ref{tab:4} in the appendix.

\subsection{Three-parameter residual model}

We try to perform comparative experiments between the translational model and the symmetrical model to explore the presence of symmetry in the pulse shapes of the fast-varying component. After some attempts, we have found that it is difficult to define the translational model with only two parameters. Once the residual of a GRB light curve are obtained, the $t_{mirror}$ can divide the residual into two parts. For the symmetrical model, the tail of the left residual is automatically aligned with the head of the right residual. However, for the translational model, what we can do is to move the left part until $t_{start}$ and $t_{mirror}$ coincide, which does not work for residuals with a long smooth head. Therefore, we introduce a three-parameter model to solve this problem. The three parameters are splitting time $t$, translation parameter $\Delta t$, and stretching parameter $s$. The translational model is characterized as follows.

The residual are obtained using the method described in Section \ref{sec:3.1}. Subsequently, the residual are divided into two parts by selecting a splitting time $t$, and the left residual are translated by a translation parameter $\Delta t$ and stretched by a stretching parameter $s$ until the maximum value of the CCF is achieved. It is important to note that we adopt a new form of CCF with a no-mean subtracted definition, expressed as
\begin{equation}
{\rm CCF_{Band}}=\frac{\sum_i x_i y_i}{\sqrt{\sum_i x_i^2} \sqrt{\sum_i y_i^2}},
\end{equation}
which is considered more suitable for transient events such as GRBs \citep{1997ApJ...486..928B,2010ApJ...711.1073U}.

To maintain the same degrees of freedom between the symmetrical model and the translational model, a translation parameter is also introduced into the symmetrical model. The symmetrical model is characterized similarly to the translational model. The difference is that in symmetrical model, the residual on the left side needs to be folded and then align with the right side. 

A more detailed description of the steps is provided below:

\begin{enumerate}
\setlength{\parskip}{0\baselineskip}
\setlength{\itemsep}{0\baselineskip}
\item Choose a GRB pulse light curve and use iterative fitting to characterize the slow component. Subtract this component from the data to obtain the residual, which represent the fast component.

\item Cut off any residual data outside the duration window or replace it with zeros. Choose a splitting time $t$ to divide the residual into two parts.

\item Fold the left part of the residual in the symmetrical model, but not in the translational model. Translate the left part by $\Delta t$, stretch it by $s$, and calculate the ${\rm CCF_{Band}}$ with the right part. 

\item Continuously adjust the model parameters to find the maximum value of the ${\rm CCF_{Band}}$. 
\end{enumerate}

When calculating the ${\rm CCF_{Band}}$, one side of the residual light curve and the projection from the other side form one of the two signals. The ${\rm CCF_{Band}}$ of the translational and symmetrical models are indicated by ${{\rm CCF}_{tm}}$ and ${{\rm CCF}_{sm}}$, respectively. The only difference between the two models is that the symmetrical model requires a folding operation while the translational model does not, allowing for a direct comparison between them. If there is also symmetry exist in the pulse shapes, we expect the symmetrical model to outperform the translational model, indicating that ${{\rm CCF}_{sm}> {\rm CCF}_{tm}}$ for the majority of GRBs.

To avoid the huge calculation resulting from traversing parameters, we adopted a global optimization algorithm known as Differential Evolution \citep{storn1997differential}. The differential evolution is a heuristic algorithm based on population evolution, similar to genetic algorithm. It has a simple structure, fast convergence, and strong robustness, and is commonly used for finding global optimal solutions for optimization problems characterized by nonlinear, multimodal, and high-dimensional relationships \citep{das2010differential}. In this work, we refer to the publicly available code of \emph{scikit-opt} \footnote{\url{https://github.com/guofei9987/scikit-opt}} to fit the residuals.

\section{Results} \label{sec:4}
The three-parameter residual model is applied to the BATSE data. The selected data are the same as \cite{2021ApJ...919...37H}. To avoid bias of human-eye, we removed the ``multi-pulse" GRBs. To avoid the inconsistency of pulse definition, as well as the instrument selection effect, we do not include GBM data here. The resulting CCF values and parameter estimates for the symmetrical and translational models are listed in Table \ref{tab:4}. The subscripts ``$sm$'' and ``$tm$'' are used to denote the symmetrical and translational models, respectively.

\begin{figure}[t!]
\epsscale{1.08}
\plottwo{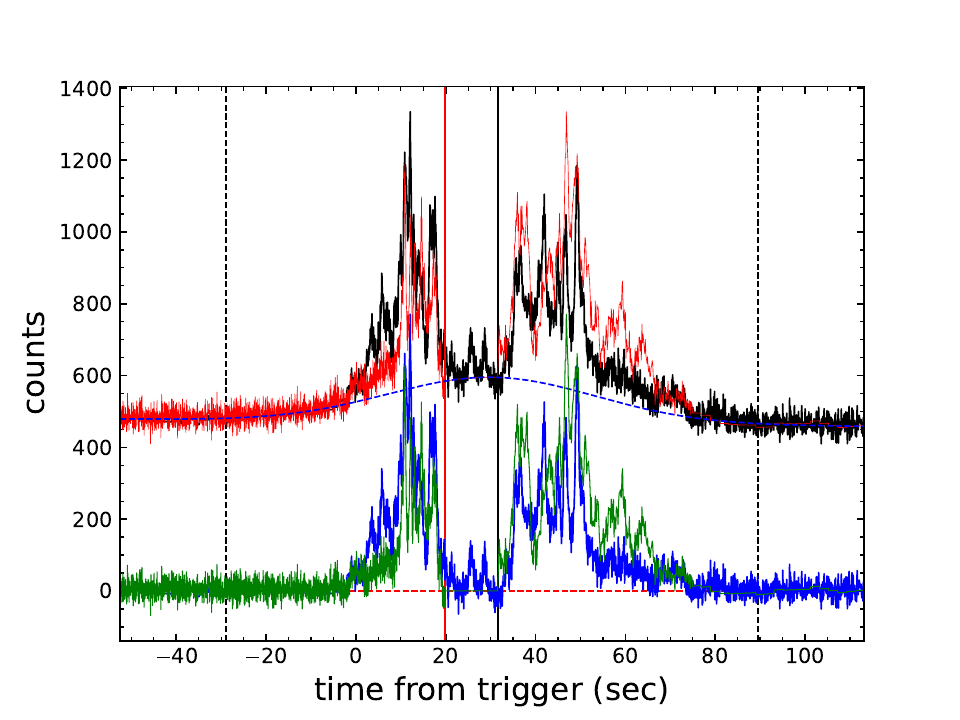}{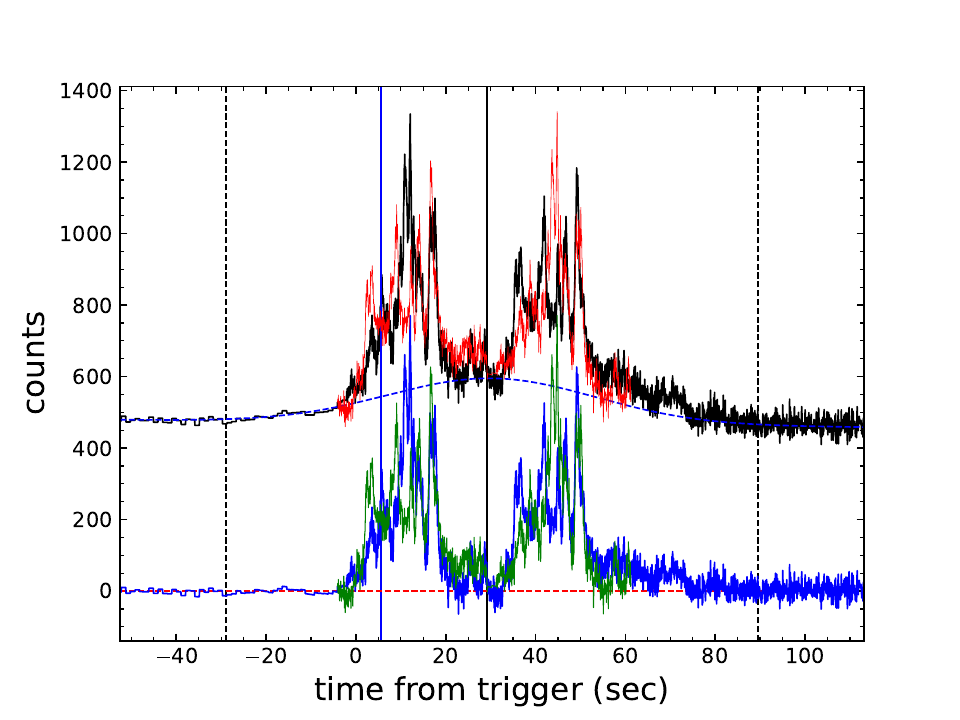}
\plottwo{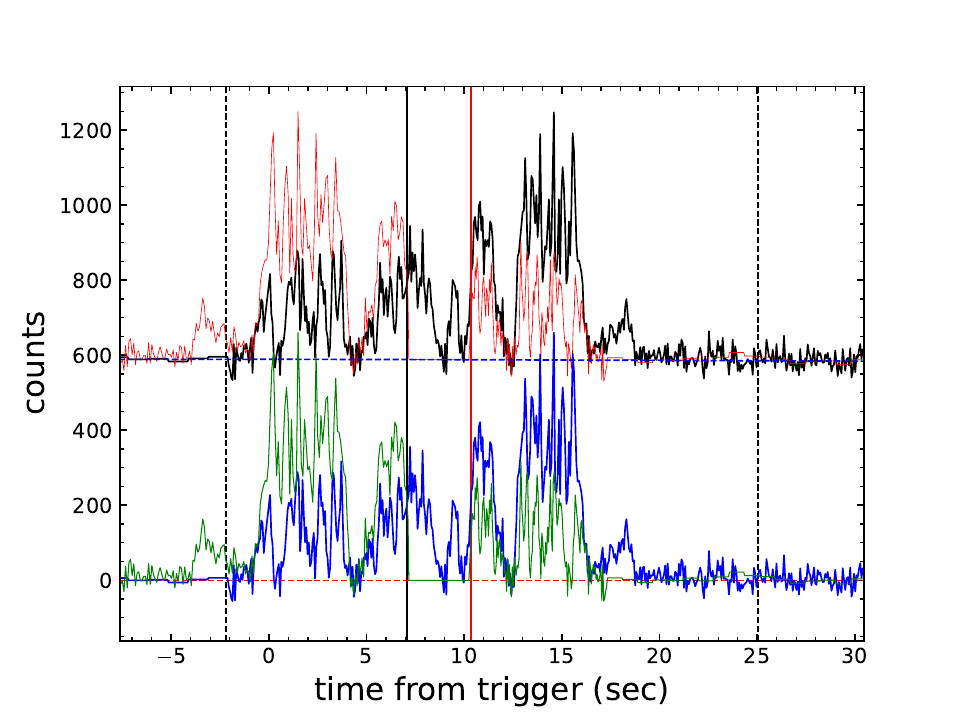}{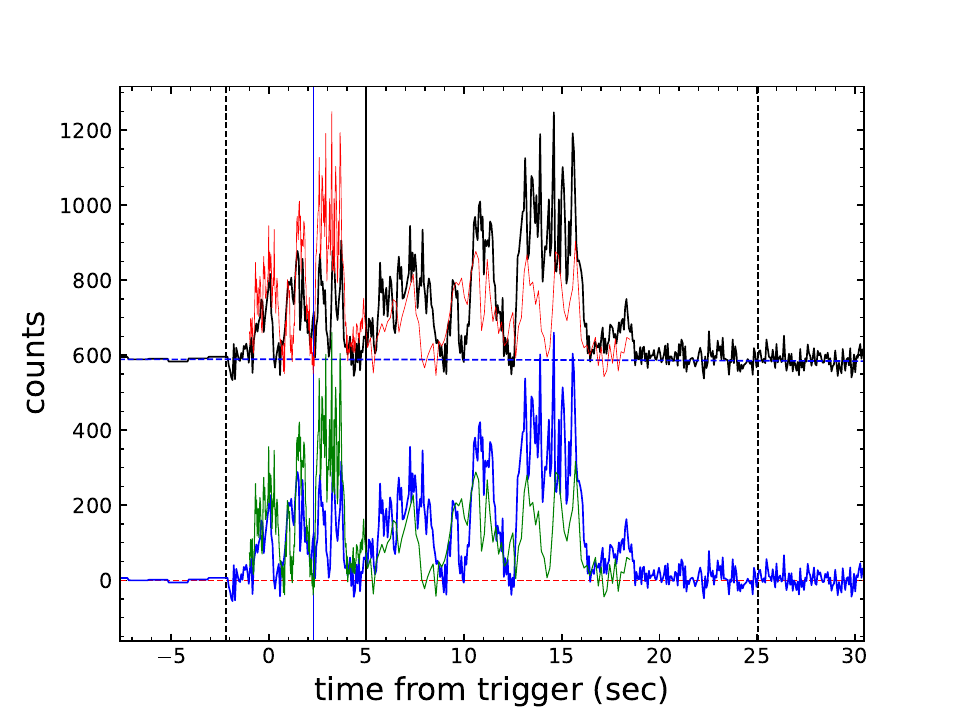}
\plottwo{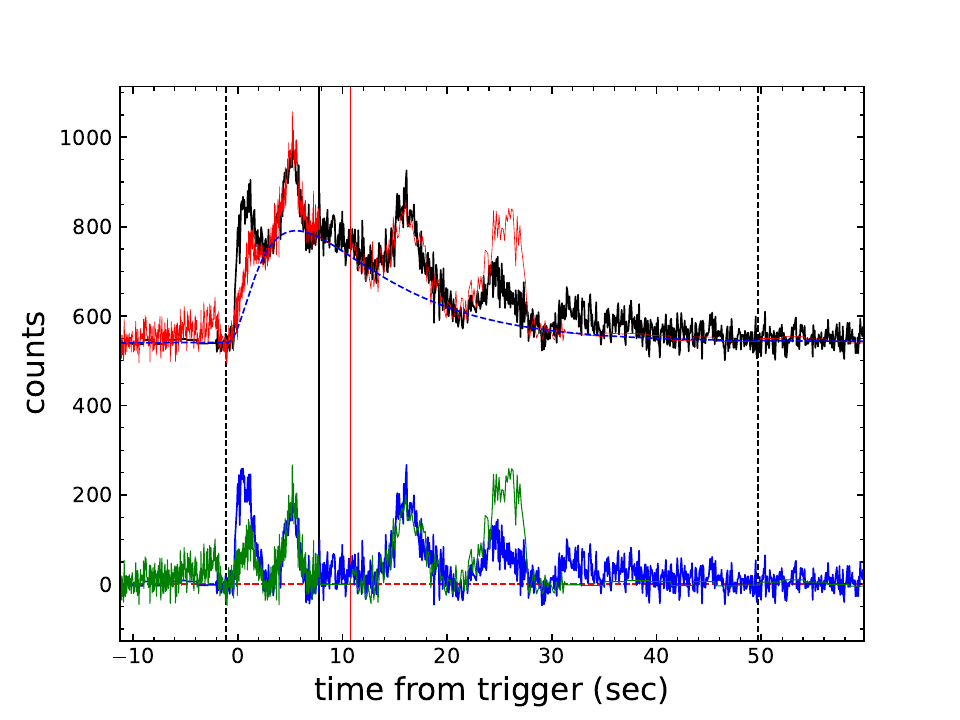}{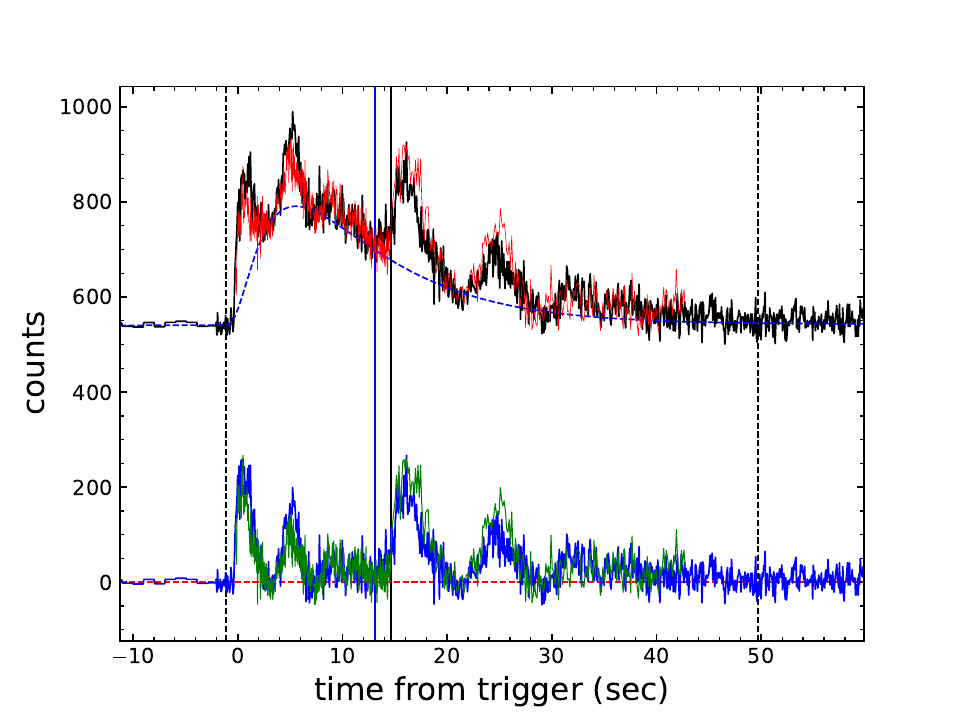}
\caption{Three-parameter residual model applies on BATSE pulse 130, 160 and 548. The figure displays symmetrical and translational models for analyzing the three pulses, along with a blue dashed line representing the identified monotonic/slow component, a vertical dashed line indicating the pulse duration window, a black vertical line indicating the splitting time, and a red dashed line at the zero horizontal line. In the symmetrical model, the tail of the left residual ($t_{sm}$) moves to the red vertical line, while in the translational model, the head of the left part ($t_{start}$) moves to the blue vertical line.
\label{fig:4}}
\end{figure}

Some visual results are presented in Figure \ref{fig:4}. 
The iterative fitting procedure is used to identify the slow component, and the fast component is obtained by subtracting it from the original data. The residual is then divided into two parts using the splitting time, and one part is translated and stretched to align with the other part. The green curve in the figure represents the fitting to the fast component, and it is added to the slow component to obtain the fitting of the whole light curve, which is represented by the red curve. We take BATSE pulses 130, 160, and 548 as examples since they all fit well with the original temporally symmetric model. In three-parameter residual model, all of them exhibit a high value of ${\rm CCF}_{sm}$, but as well as of ${\rm CCF}_{tm}$. The residual of BATSE pulse 130 exhibit two main peaks and the CCF results show that it is more suitable for the symmetrical model with ${\rm CCF}_{tm}=0.837$ and ${\rm CCF}_{sm}=0.901$. For BATSE pulse 160, iterative fitting shows that there is almost no slow component. This is a normal phenomenon. If the fast-varying and slow-varying components represent two independent radiation processes, it is possible for the slow-varying component to be too weak to be detected by the instrument, resulting in only the fast-varying component being observed. The maximum value of two models is almost equal with ${\rm CCF}_{tm}=0.832$ and ${\rm CCF}_{sm}=0.845$. However, in the symmetrical model, the light curve on each side is recognized as two peaks, while in the translational model, the light curve on one side appears to have three peaks. For BATSE pulse 548, there are four clear peaks in the residual, and the CCF results are ${\rm CCF}_{tm}=0.871$ and ${\rm CCF}_{sm}=0.739$, indicating that it is more suitable for the translational model. 

\begin{figure}[t!]
\epsscale{1.05} \plotone{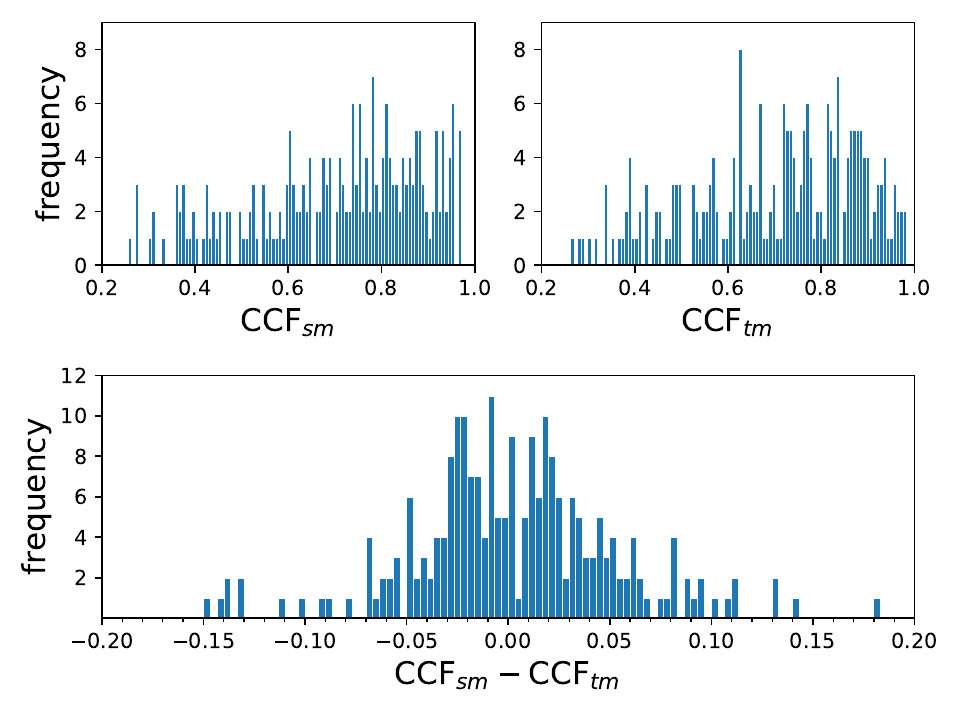}
\caption{The distributions of the results for 226 BATSE GRBs. The figure displays the distribution of ${\rm CCF}_{tm}$ for the translational model (upper left panel), ${\rm CCF}_{sm}$ for the symmetrical model (upper right panel), and the distribution of the difference in CCF between the two models (lower panel).}
\label{fig:5}
\end{figure}

The distribution of CCF values for 226 BATSE GRBs is presented in Figure \ref{fig:5}. For each GRB, the parameter results and $\rm CCF_{Band}$ values of the two models are listed in Table \ref{tab:3} in the appendix. The distribution of CCF values for both models showed no significant differences. We aim to investigate whether the results of the two models exhibit significant differences for a given GRB light curve. To achieve this, we calculate the difference between the ${\rm CCF}_{sm}$ and ${\rm CCF}_{tm}$ values for each GRB and display them as a column diagram in the lower panel of Figure \ref{fig:5}. If one of the models is more appropriate for a given GRB, we expect to observe a distribution that deviates significantly from zero. However, our analysis reveals that the difference between the two models is negligible, with the results following a Gaussian distribution centered at zero. Out of the 226 BATSE GRB pulses analyzed, 114 show ${\rm CCF}_{sm}>{\rm CCF}_{tm}$ while 112 show ${\rm CCF}_{sm}<{\rm CCF}_{tm}$. Furthermore, the absolute difference between ${\rm CCF}_{sm}$ and ${\rm CCF}_{tm}$ for each GRB light curve is not greater than 0.2, indicating that the three-parameter symmetrical model does not provide a significant advantage over the three-parameter translational model. In other words, we have not found strong evidence to suggest that the pulse shapes of the fast-varying component also exhibit temporal symmetry. Approximately half of GRBs tend to show a preference for translation, while the other half tend to show a preference for symmetry.

While it is difficult to draw a universal conclusion regarding whether the residuals of most GRBs are better fit by a symmetrical or translational model, we have observed that some GRBs with distinct features are well-described by one of the two models. Specifically, BATSE pulse 109 and 121 show a high degree of consistency with the symmetrical model, while BATSE pulse 548 and 647 are well-described by the translational model. Both of the two models show the unique properties of fast components in GRB, and their mechanisms deserve further study. 

\section{Indication to the GRB scenario} \label{sec:5}

Possible explanations for temporally symmetrical residuals have been proposed in \cite{2018ApJ...863...77H} and \cite{2019ApJ...883...70H}, which mainly attribute the symmetry to the motion or material distribution of the emitting region. Additionally, it has been shown that superluminal motion may also produce a time-reversed signal \citep{2018AnP...53000333N,2020ApJ...889..122N}. \ Motivated by the reverse-forward shock model from \cite{2014ApJ...783...88H}, we suggest that symmetrical or translational signals may be produced by the following process.

\begin{figure}[t!]
\epsscale{1.0} \plotone{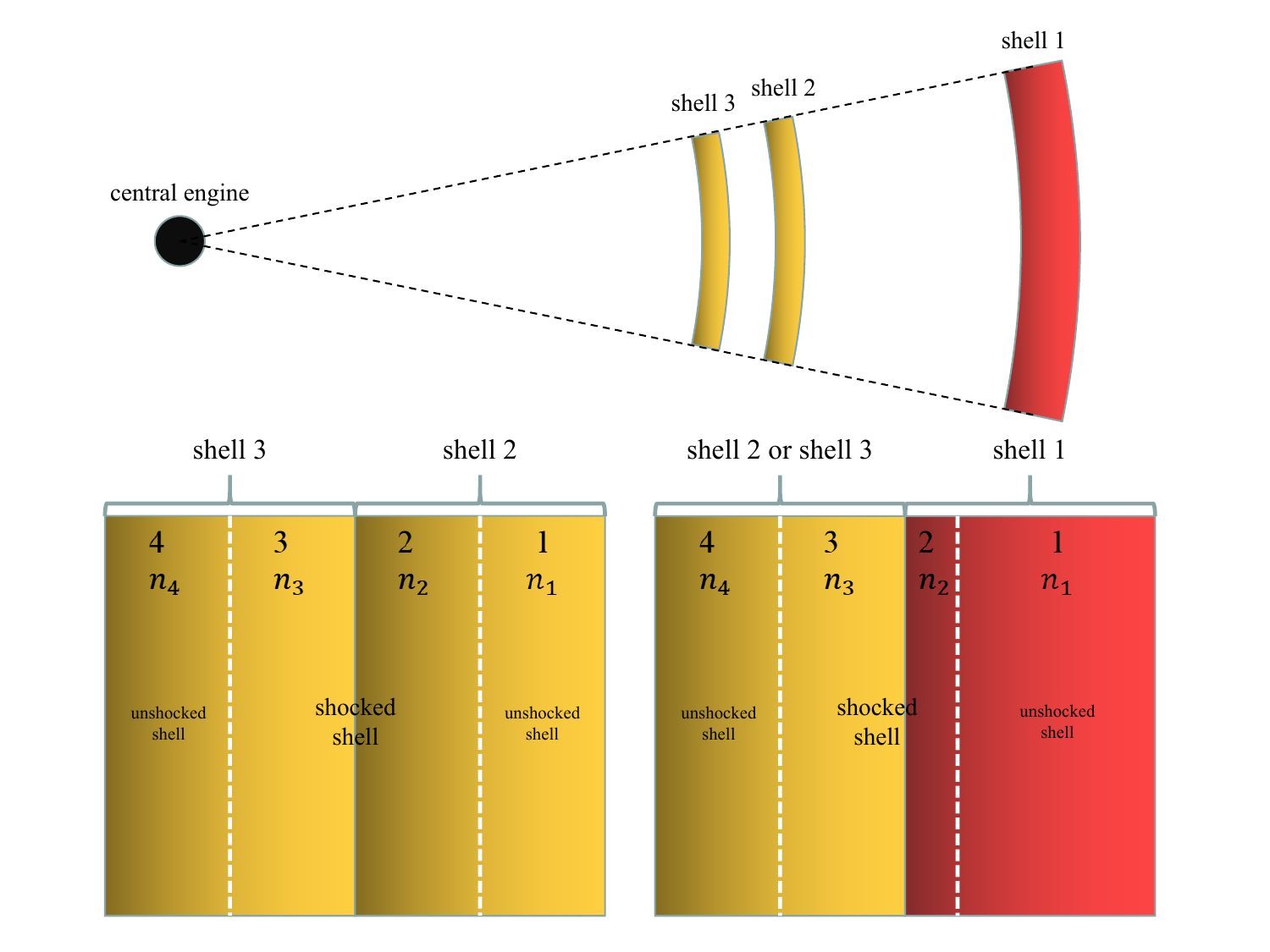}
\caption{Schematic diagram of jet structure for the symmetrical and the translational temporal structure. Assume that the fast component of the emission are from shock front (white dashed line). It varies when the shock front sweeps to denser or rarer material. The symmetrical signal can be produced by the collision between shell 2 and shell 3 with similar structures, while the translational signal cab be produced by two similar shell collision onto a bigger outer shell. 
\label{fig:6}}
\end{figure}

As shown in Figure \ref{fig:6}, intermittent activity of the central engine releases the shells. Assuming high density areas or lumps exist in both shell 2 and shell 3 with similar structures, which means the radial parameter distribution between the two shells is roughly the same. This may come from the similarity of the central engine activity for each ejection. When shell 3 catches up with shell 2, reverse-forward shocks are generated. Suppose there is extra-radiation occurring in the high-density areas or lumps when the shock front crosses. This is the most distinctive aspects compared to traditional internal shock models. In our hypothesis, the radiation from the fast-varying component originates from the shock front itself rather than the shocked material. This radiation is generated as the shock front sweeps through and disappears after passing. As the paths of the forward and reverse shock fronts are reversed, this could produce a symmetrical signal. For a time translational signal, we assume a big shell at the outermost layer, the shell 2 and shell 3 with similar structure hit shell 1 successively. A similar picture can be found at \citep{2006ApJ...646.1098Z}.
The same process of two reverse shock fronts produces a natural translational signal. It's worth noting that there may be more than two shells, as seen in the residual of BATSE pulse 2061 where three pulse structures suggest time translation.

A simple dynamic simulation was conducted without considering the radiation mechanisms. We assume that shell 2 and shell 3 have the same mass but different Lorentz factors, with $\gamma_{shell,3} = 100$ and $\gamma_{shell,2} = 50$. The Lorentz factor of the merged shell can be calculated by 
\begin{equation}
    \gamma_m = \sqrt{\frac{m_r\gamma_r+m_s\gamma_s}{m_r/\gamma_r + m_s/\gamma_s}}, 
\end{equation}
where $m_r$, $\gamma_r$ and $m_s$, $\gamma_s$ represent the mass and Lorentz factor of the fast shell and slow shell, respectively. We can also obtain the Lorentz factors of the forward shock $\gamma_{fs}$ and the reverse shock $\gamma_{rs}$ with \citep{1995ApJ...455L.143S,1997ApJ...490...92K}:
\begin{equation}
    \gamma_{fs} = \gamma_m\sqrt{(1+\frac{2\gamma_m}{\gamma_s})/(2+\frac{\gamma_m}{\gamma_s})},
\end{equation}
\begin{equation}
    \gamma_{rs} = \gamma_m\sqrt{(1+\frac{2\gamma_m}{\gamma_r})/(2+\frac{\gamma_m}{\gamma_r})}.
\end{equation}

\begin{figure}[t!]
\plottwo{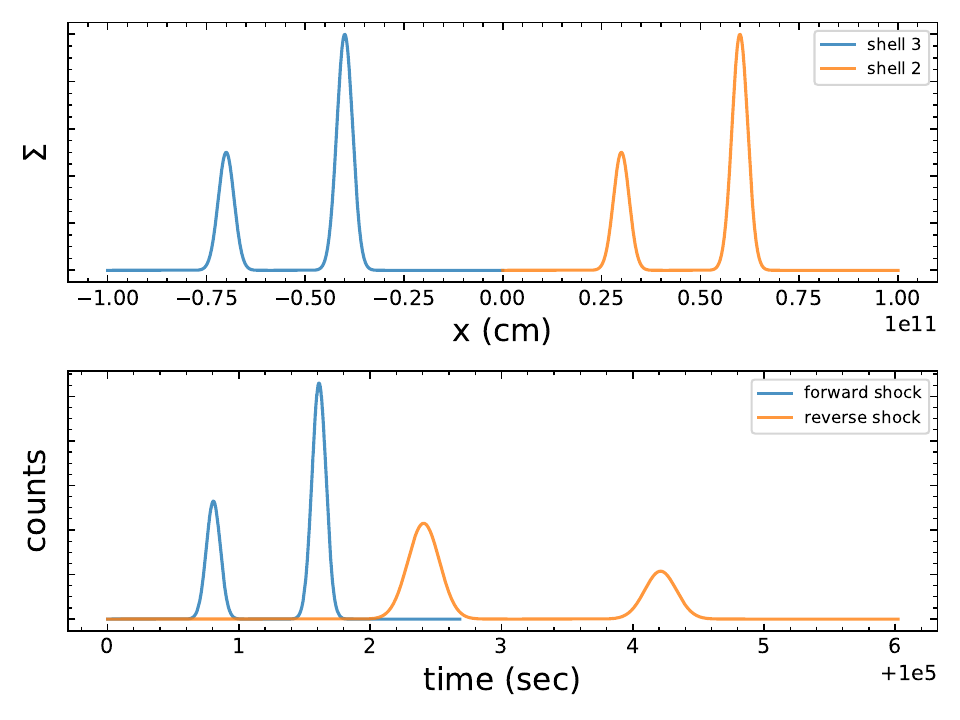}{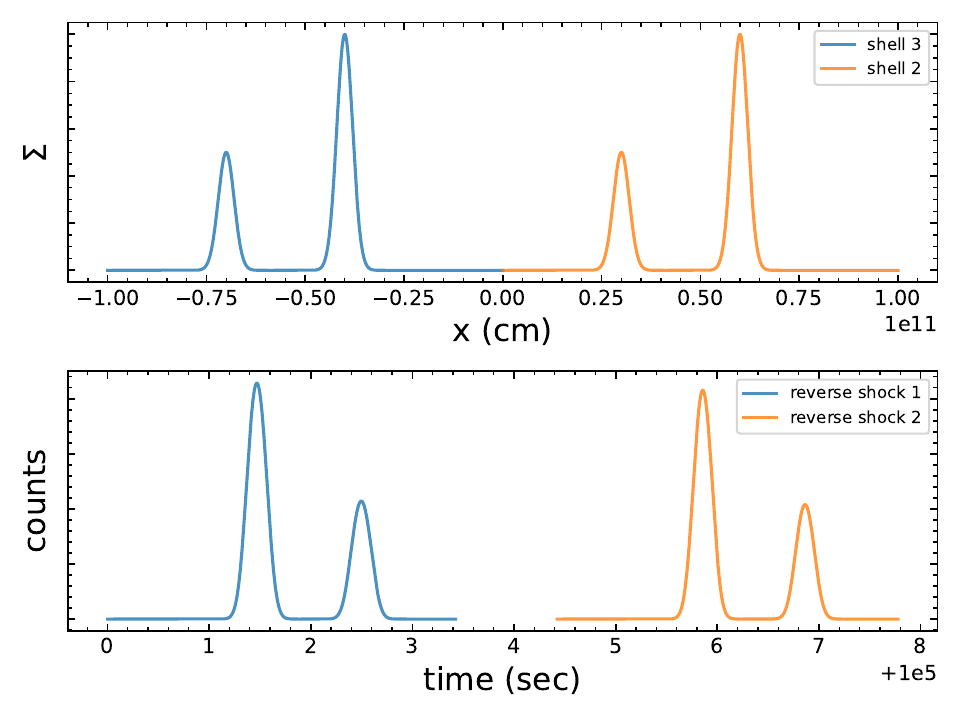}
\caption{Simulated light curve with two situation. The upper panel in each figure represents the areal density distribution of shell 2 and shell 3, the peaks represents high-density area or blocks. The left panel shows shell 2 and shell 3 collision, and the right panel simulates shell 2, shell 3 collision with shell 1 in turn.
\label{fig:7}}
\end{figure}

Figure \ref{fig:7} displays the outcomes of the simulation. Two blocks are added for each shell. When shell 2 and shell 3 interact, the relative velocity of the forward shock front is higher than the reverse shock front. Thus, after the radiation of two blocks in shell 2, the radiation of the two blocks of shell 3 begins. The latter is just the reverse version of the former with time reversal and stretching. For the collision between shell 2 and shell 1, and shell 3 and shell 1 in turn, we set $m_{shell,1} = 100\ m_{shell,2}$ and $\gamma_{shell,1} = 10$. The reverse shock dominates the radiation, and the light curve of the two collisions shows a version of time translation. Though in this model, it only needs two or three shells ejected from the central engine, it is not a restriction for all GRBs. In \cite{2021ApJ...919...37H}, there are a small fraction of the sample should be taken as multiple structures, while we simply omitted those samples in this work. For those GRBs, the central engine should have ejected more shells. Although the actual light curve is more complex, our model can easily explains these features. For a more complex time-reversed structure, we attribute these characteristics to the inherent complexity of the shell structure, such as the number of the higher density regions in the shell, the values of densities, the micro-physics parameters, etc. Too many free parameters makes the exact light curve fitting not meaningful. In Figure \ref{fig:7}, we only assumed the presence of two high-density regions within each shell, but in reality, there may be more regions of varying sizes. This leads to the fast-varying features observed in the light curve. For each structure, the pulse order before the reflection time is exactly opposite to the pulse order after the reflection. It is important to note that the pulse order is determined based on the pulse amplitude. In our model, this is related to the region density and shock front velocity. Additionally, the ratio of pulse duration before and after the reflection time, denoted as $s_{mirror}$ in the original model, is now associated with the ratio of velocities of the forward and reverse shock fronts in the co-moving frame in our model. 

We also notice that \cite{2014ApJ...783...88H} and \cite{2018ApJ...863...77H} mentioned the reverse-forward shock model cannot easily reproduce the multiple pulses. The reason is that the dips of lower density could be smeared out, as shown in the simulations of \cite{2004ApJ...611.1021K}. We argue that our model is phenomenological, which does not solely depend on the density fluctuation. It could be other structures, which induce the fast pulses. Indeed, the actual cause for the emission of the fast component still needs further investigation.

This scenario could be identified or denied by the future observations, especially when the full Square Kilometre Array (SKA) starts to observe \citep{2009IEEEP..97.1482D}. The symmetrical scenario produces smooth emission from collision of two shells with higher Lorentz factor, while the translational scenario produces smooth emission from successive collision onto the slow outermost shell. This makes the difference of two scenarios, that for the symmetrical scenario the Lorentz factor for the emitting region is higher comparing to the translational scenario. The radio emission could come together with the prompt $\gamma$-ray emission. However, the radius at prompt phase is small enough to be optical thick for the radio band. Therefore, the Lorentz factor acts as an important role for the radio emission intensity. We predict for the symmetrical scenario, the prompt radio emission is much higher than that for the translational scenario. On the other hand, shell 1 contains more material. It makes the radio emission from the shocked shell 1 lasts longer time. In conclusion, the GRBs with symmetrical structure should have stronger prompt radio emission, while the GRBs with translational structure should be weaker but last longer. The detailed estimation of the prompt radio flux could refer to \cite{2006ApJ...646.1098Z} (see eqs. (12-13) and (25-28)). As estimating from 1st term in eq. (12), for the translational case, the 1 GHz peak flux could be around $2.4 \times 10^{-10} {\rm Jy}$ for a source being at $10^{28}$ cm. For the symmetrical case, the collision radius should be far as the Lorentz factors are higher, and we turn to the 1st one in eq. (27). The estimated 1 GHz peak flux could be around $3.5 \times 10^{-6} {\rm Jy}$, if we set the Lorentz factor of shell 1 being $100$. There should be more complex when the detailed parameters combination are considered.

\section{Conclusion and discussion} \label{sec:6}

In this study, we firstly verified the result presented in \cite{2021ApJ...919...37H} that the majority of GRB pulse light curves can be characterized by a smooth single-peaked component and a complex residual structure which is temporally symmetric. Our analysis utilized data from BATSE which replicated the previous results. Then we extended to the GBM data and got similar results. We found that the results obtained from the BATSE data showed a high ratio of success, with $85\%$ of the GRBs fitting the model well, which is in line with the findings in \cite{2021ApJ...919...37H} where a ratio of $86.6\%$ was reported. When we applied the model to GBM 64 ms data, we obtained a lower ratio of $73.6\%$ and $77.1\%$, which could be attributed to the different effective areas between BATSE and GBM. This leads to varying signal-to-noise ratios that may impact the criteria of the model. Since the original model can only identify the symmetry of pulse orders but not the symmetry of pulse shapes, in order to further confirm this symmetry, we have designed a new model that can validate the symmetry of pulse shapes. However, our comparison between the translational model and the symmetrical model did not yield strong evidence to suggest that the symmetrical model is better. The calculated value of $\rm CCF_{Band}$ was almost the same between the two models. As shown in Figure \ref{fig:5}, about half can be considered symmetrical and another half can be considered as translational. Both features could come from the structure of the ejected shells, i.e., the fast component represents the structure. We assume there are two shells with similar structure, if they collide each other, the fast component is symmetrical, while if they collide to the external shocked shell one by one, the fast component is translational. We suggest the future full SKA could be able to test this scenario because of its large field of view as well as high sensitivity. The Five-hundred-meter Aperture Spherical radio Telescope (FAST) also has a very high sensitivity \citep{2011IJMPD..20..989N}. However, the small field of view makes the detection rate very low for the prompt radio emission.

We noticed that in the original model, the uncertainty of parameter $s_{mirror}$ is the main criterion rather than the value of CCF. We decided to abandon this approach as the uncertainty in the three-parameter model would increase, rendering the original criterion invalid. We also notice that the CCF value of different GRBs cannot be used for direct comparison, for the GRBs with bright residual usually show a larger CCF value. However, within our model, it is feasible to compare the translational and symmetrical models for a same GRB light curve without such concerns.

It is worth mentioning that, relative to the new method for identifying monotonic components, the original method just introduces a systematic deviation with convex shape, resulting in residuals with a concave systematic deviation. This is the reason why the residuals obtained by direct fitting show components that are obviously less than zero in the pulse duration window but near zero outside of the window. This deviation favors the symmetrical model but is against the translational model, which is the main reason why we developed the new approach to subtract the monotonic component.

Notice that the distribution of ${\rm CCF}_{sm}-{\rm CCF}_{tm}$ concentrates around 0. Therefore, only a small part of the fast component light curves should be determinately classified as symmetrical or translational. At present, it remains challenging to definitively categorize GRBs into distinct groups of translation and symmetry. We expect more light curves from various GRB telescopes could be considered in the further investigation.

\ \

We thank Weihua Lei, Yu Liu, Zipei Zhu and Kai Wang for helpful discussions.
This work is in part supported  by the National Key R\&D Program of China (2022SKA0130100), and by the National Natural Science Foundation of China (Grant Nos. 12041306 and U1931203). We also acknowledge the science research grants from the China Manned Space Project with No. CMS-CSST-2021-B11. The computation is completed in the HPC Platform of Huazhong University of Science and Technology.

\bibliography{ref}{}
\bibliographystyle{aasjournal}

\clearpage
\appendix
\setlength{\tabcolsep}{1.5pt}
\begin{longtable}{l c c c c c c c c c c c c c c c}
\caption{Temporally symmetric model applies to BATSE dataset\label{tab:2}} \\
\hline
Pulse & $B_0$ & $BS$ & $A$ & $t_s$ & $\tau_1$ & $\tau_2$ & $t_0$ & $\sigma$ & CCF & $R$ & $p$ & $t_{mirror}$ & $\sigma_{t,mirror}$ & $s_{mirror}$ & $\sigma_{s,mirror}$\\ \hline
\endfirsthead

\caption{Continued} \\
\hline
Pulse & $B_0$ & $BS$ & $A$ & $t_s$ & $\tau_1$ & $\tau_2$ & $t_0$ & $\sigma$ & CCF & $R$ & $p$ & $t_{mirror}$ & $\sigma_{t,mirror}$ & $s_{mirror}$ & $\sigma_{s,mirror}$\\ \hline
\endhead

\hline
\multicolumn{3}{l}{\textit{Continued on next page}} \\
\endfoot

\hline
\endlastfoot

  105 &  619.9 &  0.209 &   2763.0 &    1.93 &       10.75 &    0.26 &     ... &     ... & 0.812 &  5.97 &     0.00 &      4.16 &     0.27 &      0.19 &     0.14 \\
  107 &  547.1 &  0.023 &     37.0 &     ... &         ... &     ... & -150.29 &   41.97 & 0.446 &  0.45 &     1.00 &   -186.94 &    23.38 &      0.88 &     0.41 \\
  109 &  643.3 &  0.087 &    540.8 &   -0.28 &        4.37 &   36.98 &     ... &     ... & 0.575 &  4.40 &     0.00 &     49.40 &     0.23 &      0.70 &     0.01 \\
  111 &  541.3 &  0.003 &     86.5 &  -16.83 &       16.15 &    26.5 &     ... &     ... & 0.180 &  1.04 &     0.42 &      1.08 &    20.49 &      0.21 &     0.56 \\
  114 &  550.0 & -0.082 &    118.2 &     ... &         ... &     ... &   -0.43 &    2.25 & 0.582 &  1.45 &     0.00 &     -0.11 &     0.85 &      2.37 &     0.77 \\
  121 &  480.5 &  0.079 &    115.4 &     ... &         ... &     ... &   30.22 &   35.29 & 0.654 &  2.23 &     0.00 &     23.70 &     3.44 &      0.61 &     0.23 \\
  130 &  474.9 & -0.106 &    316.9 &     ... &         ... &     ... &    30.3 &   24.15 & 0.678 &  5.24 &     0.00 &     27.04 &     0.11 &      1.07 &     0.01 \\
  160 &  587.8 &  0.004 &    239.3 &     ... &         ... &     ... &   11.43 &    5.53 & 0.595 &  4.65 &     0.00 &     14.21 &     1.51 &      0.84 &     0.39 \\
\end{longtable}

\setlength{\tabcolsep}{1.6pt}
\begin{longtable}{l c c c c c c c c c c c c c c c}
\caption{Temporally symmetric model applies to GBM dataset\label{tab:3}} \\
\hline
Trigger & $B_0$ & $BS$ & $A$ & $t_s$ & $\tau_1$ & $\tau_2$ & $t_0$ & $\sigma$ & CCF & $R$ & $p$ & $t_{mirror}$ & $\sigma_{t,mirror}$ & $s_{mirror}$ & $\sigma_{s,mirror}$\\ \hline
\endfirsthead

\caption{Continued} \\
\hline
Trigger & $B_0$ & $BS$ & $A$ & $t_s$ & $\tau_1$ & $\tau_2$ & $t_0$ & $\sigma$ & CCF & $R$ & $p$ & $t_{mirror}$ & $\sigma_{t,mirror}$ & $s_{mirror}$ & $\sigma_{s,mirror}$\\ \hline
\endhead
\hline
\multicolumn{3}{l}{\textit{Continued on next page}} \\
\endfoot

\hline
\endlastfoot
bn200101861 & 150.9 & -0.002 & 1388.7 &     ... &     ... &       ... &    11.9 &    0.77 & 0.653 &  9.00 &     0.00 &     11.65 &     0.11 &      2.49 &     0.48 \\
bn200103689 & 146.0 &  0.017 &    7.9 &   -9.14 &    0.46 &    165.16 &     ... &     ... & 0.071 &  1.04 &     0.00 &    136.79 &   108.88 &      1.80 &     0.32 \\
bn200111633 & 283.6 & -0.016 &  331.3 &    3.24 &   49.38 &      0.09 &     ... &     ... & 0.810 &  2.75 &     0.00 &      5.37 &     0.11 &      1.00 &     0.16 \\
bn200125864 & 118.8 &  0.014 & 1005.1 &    0.95 &    0.21 &      2.53 &     ... &     ... & 0.748 & 18.24 &     0.00 &      2.19 &     0.05 &      0.25 &     0.02 \\
bn200130417 & 134.2 &  0.020 &  516.1 &    -0.5 &     2.2 &      0.71 &     ... &     ... & 0.841 &  3.62 &     0.00 &      1.47 &     0.26 &      0.43 &     0.08 \\
bn200219998 & 190.0 & -0.051 &  306.4 &    -5.2 &   75.37 &      0.91 &     ... &     ... & 0.672 &  3.24 &     0.00 &      4.23 &     0.61 &      0.63 &     0.45 \\
bn200227306 & 145.4 & -0.022 &   91.6 &  -27.21 &  661.73 &      2.07 &     ... &     ... & 0.341 &  1.25 &     0.00 &      3.96 &    51.67 &      0.86 &     0.36 \\
bn200228291 & 162.6 & -0.016 &  279.1 &   -4.01 &   90.58 &      0.35 &     ... &     ... & 0.739 &  3.25 &     0.00 &      1.07 &     0.14 &      0.85 &     0.20 \\
\end{longtable}

\setlength{\tabcolsep}{1.5pt}
\begin{longtable}{l c c c c c c c c c c c c c c c c}
\caption{The there-parameter models apply to BATSE dataset\label{tab:4}} \\
\hline
\multirow{2}{*}{Pulse} & \multirow{2}{*}{$B_0$} & \multirow{2}{*}{$BS$} & \multirow{2}{*}{$A$} & \multirow{2}{*}{$t_s$} & \multirow{2}{*}{$\tau_1$} & \multirow{2}{*}{$\tau_2$} & \multirow{2}{*}{$t_0$} & \multirow{2}{*}{$\sigma$} & 
\multicolumn{4}{c}{Symmetrical model} &
\multicolumn{4}{c}{Translational model} \\
\cline{10-13}
\cline{14-17}
\multicolumn{9}{c}{} & $t_{sm}$ & $s_{sm}$ & $\Delta t_{sm}$ & ${\rm CCF}_{sm}$ & $t_{tm}$ & $s_{tm}$ & $\Delta t_{tm}$ & ${\rm CCF}_{tm}$ \\ \hline
\endfirsthead

\caption{Continued} \\
\hline
\multirow{2}{*}{Pulse} & \multirow{2}{*}{$B_0$} & \multirow{2}{*}{$BS$} & \multirow{2}{*}{$A$} & \multirow{2}{*}{$t_s$} & \multirow{2}{*}{$\tau_1$} & \multirow{2}{*}{$\tau_2$} & \multirow{2}{*}{$t_0$} & \multirow{2}{*}{$\sigma$} & 
\multicolumn{4}{c}{Symmetrical model} &
\multicolumn{4}{c}{Translational model} \\
\cline{10-13}
\cline{14-17}
\multicolumn{9}{c}{} & $t_{sm}$ & $s_{sm}$ & $\Delta t_{sm}$ & ${\rm CCF}_{sm}$ & $t_{tm}$ & $s_{tm}$ & $\Delta t_{tm}$ & ${\rm CCF}_{tm}$ \\ \hline
\endhead
\hline
\multicolumn{3}{l}{\textit{Continued on next page}} \\
\endfoot

\hline
\multicolumn{17}{p{1\linewidth}}{\textbf{Notes.} For each GRB pulse, the monotonic/slow component identified by the symmetrical model and the translational model is identical. The difference lies in the three parameters used by the two models to characterize the residual/fast component: the splitting time $t$, the stretching parameter $s$, and the translation parameter $\Delta t$. Additionally, the quality of each model is evaluated using the $\rm CCF_{Band}$. The parameters with subscripts ``$sm$" and ``$tm$" are used to represent the symmetrical and translational models, respectively.
} \\
\endlastfoot
  105 &  592.4 & -0.104 &   2647.6 &     2.0 &        10.35 &     0.24 &     ... &     ... &    4.32 & 0.47 &    -0.26 &  0.920 &    4.35 & 1.06 &     1.11 &  0.817 \\
  107 &  539.2 &  0.002 &     38.5 &     ... &          ... &      ... &  -144.3 &   51.47 & -150.20 & 1.41 &     0.03 &  0.471 & -139.31 & 1.51 &    84.61 &  0.495 \\
  109 &  630.5 &  0.106 &    489.3 &    1.21 &        4.032 &   23.249 &     ... &     ... &   38.81 & 0.67 &    25.22 &  0.782 &   57.76 & 0.58 &    42.15 &  0.718 \\
  111 &  538.3 &  0.005 &     73.8 &  -50.44 &      298.478 &    10.92 &     ... &     ... &   10.66 & 0.19 &    -3.15 &  0.442 &   16.99 & 0.59 &    38.76 &  0.451 \\
  114 &  543.0 & -0.106 &     82.6 &     ... &          ... &      ... &   -0.66 &    2.41 &   -1.52 & 0.27 &    -0.12 &  0.731 &   -1.27 & 0.88 &     2.48 &  0.739 \\
  121 &  479.9 &  0.073 &     62.0 &     ... &          ... &      ... &    20.5 &   33.34 &   16.86 & 0.58 &    17.07 &  0.806 &   18.77 & 1.20 &    61.85 &  0.666 \\
  130 &  472.5 & -0.124 &    126.5 &     ... &          ... &      ... &    30.2 &   23.18 &   31.60 & 0.51 &   -23.33 &  0.901 &   29.26 & 1.05 &    34.49 &  0.837 \\
  160 &  582.0 &  0.006 &      7.4 &     ... &          ... &      ... &   -2.18 &   21.15 &    7.08 & 1.31 &     3.26 &  0.845 &    4.96 & 0.44 &     4.46 &  0.832 \\
\end{longtable}
\end{document}